\documentclass[aps,prx,reprint,superscriptaddress,showpacs,longbibliography]{revtex4-1}
\usepackage{braket}
\usepackage{amsmath}
\usepackage{amsfonts}
\usepackage{amssymb}
\usepackage{graphicx}
\usepackage{color}
\usepackage{dcolumn}
\usepackage{bm}
\usepackage{standalone}
\usepackage{physics}

\begin{document}
\title{Deep reinforcement learning for robust quantum optimization}

\author{Vegard B. S{\o}rdal}
\email[]{vegardbs@fys.uio.no}
\affiliation{Department of Physics, University of Oslo, 0316 Oslo, Norway}

\author{Joakim~Bergli}
\affiliation{Department of Physics, University of Oslo, 0316 Oslo, Norway}

\date{\today}

\begin{abstract}
Machine learning techniques based on artificial neural networks have been successfully applied to solve many problems in science. One of the most interesting domains of machine learning, reinforcement learning, has natural applicability for optimization problems in physics. In this work we use deep reinforcement learning and Chopped Random Basis optimization, to solve an optimization problem based on the insertion of an off-center barrier in a quantum Szilard engine. We show that using designed protocols for the time dependence of the barrier strength, we can achieve an equal splitting of the wave function (1/2 probability to find the particle on either side of the barrier) even for an asymmetric Szilard engine in such a way that no information is lost when measuring which side the particle is found. This implies that the asymmetric non-adiabatic Szilard engine can operate with the same efficiency as the traditional Szilard engine, with adiabatic insertion of a central barrier. We compare the two optimization methods, and demonstrate the advantage of reinforcement learning when it comes to constructing robust and noise-resistant protocols.\\
\end{abstract}

\maketitle
	
\section{Introduction}
Machine Learning is becoming an essential tool for data analysis and optimization in a wide variety of scientific fields, from molecular \cite{butler2018machine} and medical science \cite{shen2017deep} to  astronomy \cite{primack2018deep}. One of the most exiting development in machine learning, comes from combining reinforcement learning \cite{bellman1957markovian} with deep neural networks \cite{mnih2015humanlevel}. Reinforcement Learning (RL) differs from supervised and unsupervised learning and is based on letting an agent learn how to behave in a desired way by taking actions in an environment and observing the effect of the action on the environment. In order to define the optimal behavior of the agent, we give it feedback in the form  of a reward based on the effect of its previous action. If the action changes the environment into a more desirable state we give it a positive reward, while if it had negative consequences we give it a negative reward. Recently RL has enjoyed increasing popularity in quantum physics, and have been used to explore the quantum speed limit \cite{bukov2018reinforcement,zhang2018automatic_RL}, protect qubit systems from noise \cite{fosel2018reinforcement_RL}, design new photonic experiments \cite{melnikov2018active_RL}, and many other applications \cite{biamonte2017quantum_RL,dong2008quantum_RL,lamata2017basic_RL}.
For an excellent review of the application of machine learning in physics, see \cite{mehta2018high_RL_review}.\\

We use deep reinforcement learning (DRL), specifically Deep-Q Learning (DQL) \cite{mnih2015humanlevel} and Deep Deterministic Policy Gradient (DDPG) \cite{lillicrap2015continuous}, to solve an optimization problem based on the barrier insertion of an asymmetric (off-center insertion) quantum Szilard engine, which we will motivate it the following paragraphs. The goal is to find barrier insertion protocols that effectively achieves equal splitting of the wave function of a single-particle-box. We compare the results from DRL with those obtained by using chopped random basis optimization \cite{PhysRevA.84.022326}, a more traditional optimization algorithm. Finally, since it can be difficult to experimentally determine the exact asymmetry, we show that DRL can be used to find robust protocols, which performs well for a range of asymmetries. We do this by simultaneously training on many instances of the single-particle-box (SPB), where each instance has a different asymmetry. This is essentially the same as training in an environment with a noisy Hamiltonian, as in \cite{fosel2018reinforcement_RL,zhang2018automatic_RL}.\\

The Szilard engine is a classic example of a information processing system, which can convert one bit of Shannon information (obtained by a binary measurement) into an amount $ k_BT\ln 2 $ of useful work \cite{Szilard1929}. This is done by inserting a barrier in the center of a SPB, performing a measurement to determine which side of the barrier the particle is found (giving one bit of Shannon information), and then letting the compartment the particle occupies isothermally expand into the empty one resulting in a work-extraction of $ k_BT\ln 2 $. This work is not free however, since the information obtained has to be stored in a memory, which subsequently has to be deleted at an energy cost of $ k_BT\ln 2 $ according to Landauer's principle \cite{landauer1961irreversibility}. Both work extraction from a Szilard engine, and Landauer's principle, have recently been experimentally confirmed \cite{toyabe2010experimental,berut2012experimental,koski2014experimental,vidrighin2016photonic}.\\

For the quantum version of the Szilard engine \cite{lloyd1997quantum}, there are some subtle differences in the entropy flow during insertion, expansion, and removal of the barrier \cite{kim2011quantum}. Moreover, the position of the particle is now described by a quantum wave function, which is divided into two parts when inserting the barrier. When adiabatically inserting a barrier in the center of a quantum SPB in its ground state, the wave function is split in half in such a way that each half becomes a new ground state in each compartment, when the barrier strength goes to infinity. The probability to find the particle on either side of the barrier after insertion becomes 1/2. However, as long as there is an asymmetry in the insertion of the barrier, i.e. it is not put exactly in the center, the adiabatic theorem guarantees that the particle will be found in the larger compartment \cite{AmJourPhys}. Since the initial state is the ground state, and the adiabatic theorem implies the time evolved state will stay in its instantaneous eigenstate, the particle always ends up in the global ground state. The global ground state is found in the larger compartment since the energy is proportional to $ L_{R(L)}^{-2} $, where $ L_{R(L)} $ is the width of the compartment on the right(left) side of the barrier.\\

If we want to achieve equal probability on both sides of the barrier for asymmetric insertion, we have to insert the barrier non-adiabatically in such a way that we excite higher eigenstates. This will in general decrease the efficiency of the quantum Szilard engine, since the measurement only determines which side the particle is found, not its exact eigenstate. However, there is one special way of obtaining exact splitting of the wave function without losing any information in the measurement, for the asymmetric Szilard engine \cite{PhysRevA.99.022121}: If we insert the barrier in such a way that the total wave function is a superposition of only the first and second eigenstate at the time of measurement, i.e. $ \ket{\Psi} = \left(\ket{\psi_1}+\ket{\psi_2}\right)/\sqrt{2} $, the which-side measurement does not result in any information loss since the second eigenstate becomes the ground state of the smaller compartment. When one now measure which compartment the particle is in, one is certain that it is in the ground state of the respective compartment.\\

Our goal is to split the wave function of a single-particle-box in the ground state, by inserting a barrier off-center, in such a way that only the second eigenstate is excited, and the probability to find the particle in all higher states are as close to zero as possible. However, finding a  protocol for the barrier insertion which will achieve this goal is non-trivial, since it will have to take advantage of complicated interference between the time-dependent eigenstates.\\

\section{Single-particle-box}
\begin{figure}[h!]
	\centering
	\includegraphics[width=0.7\linewidth]{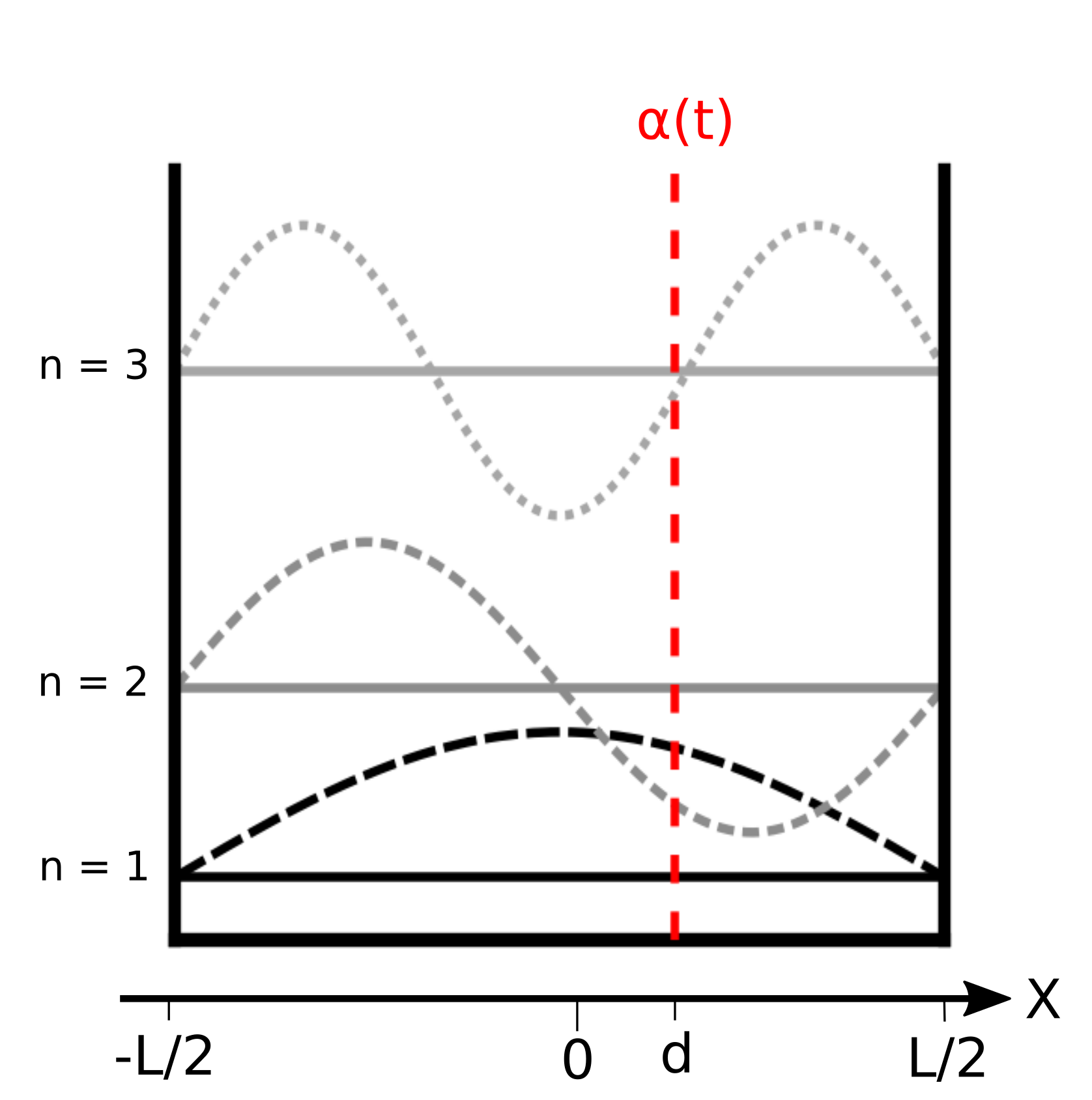}
	\caption{Illustration of a single particle box with total width $ L $. The eigenfunctions and eigenenergies are shown for the initial state $ \alpha(t)=0 $. }
	\label{fig:spb}
\end{figure}
The SPB is defined by the potential $ V(x) = 0 $ for $ x\in[-L/2,L/2] $, where $ L $ is the total width of the box, and $ V(x) = \infty $ elsewhere. The barrier is a $ \delta $-function potential inserted at $ x = d\geq0 $. An illustration of the SPB is shown in  Fig.~\ref{fig:spb}, along with its three first eigenfunctions and eigenenergies before the barrier is inserted. If $ \epsilon = 0 $ the box is split symmetrically, i.e. the width of the left and right compartment is equal. However, for $ d>0 $, the width of the left compartment becomes $ L_L = L/2 +d $, while the width of the right compartment becomes $ L_R = L/2 -d$. The time-dependent Hamiltonian of the insertion procedure is given by
\begin{equation}
\hat{H}(t) = -\frac{\hbar^2}{2m}\frac{\partial^2}{\partial x^2} +\alpha(t)\delta(x-d),
\end{equation}
where $ \alpha(t) $ is the strength of the barrier at time $ t $, and $ m $ is the mass of the particle. For the rest of this article we set $ \hbar = m = 1 $. The total wave function, $ \ket{\Psi(t)} $, can be expressed as a linear combination of the instantaneous eigenfunctions
\begin{equation}
\ket{\Psi(t)}=\sum_{n}c_n(t)\ket{\psi_n(t)}e^{i\theta_n(t)},\quad \theta_n = -\frac{1}{\hbar}\int_{0}^{t}E_n(t')~dt',
\end{equation}
where $ E_n(t) $ are the instantaneous eigenenergies when the barrier strength is $ \alpha(t) $, $ \ket{\psi_n(t)} $ are the instantaneous eigenfunctions, and $ c_n(t) $ are complex coefficients. The initial state is therefore given by $ \abs{c_1(0)}^2 = 1 $, and the goal is to construct a protocol $ \alpha(t) $, which brings us to a final state where $ \abs{c_1(T)}^2=\abs{c_2(T)}^2=1/2 $, where $ T $ is the duration of the protocol. More details on how the instantaneous eigenstates are calculated, and how the time evolution of the total wave function is numerically solved, is given in \cite{PhysRevA.99.022121}.

\section{CRAB optimization}
\begin{figure}[h!]
	\centering
	\includegraphics[width=1\linewidth]{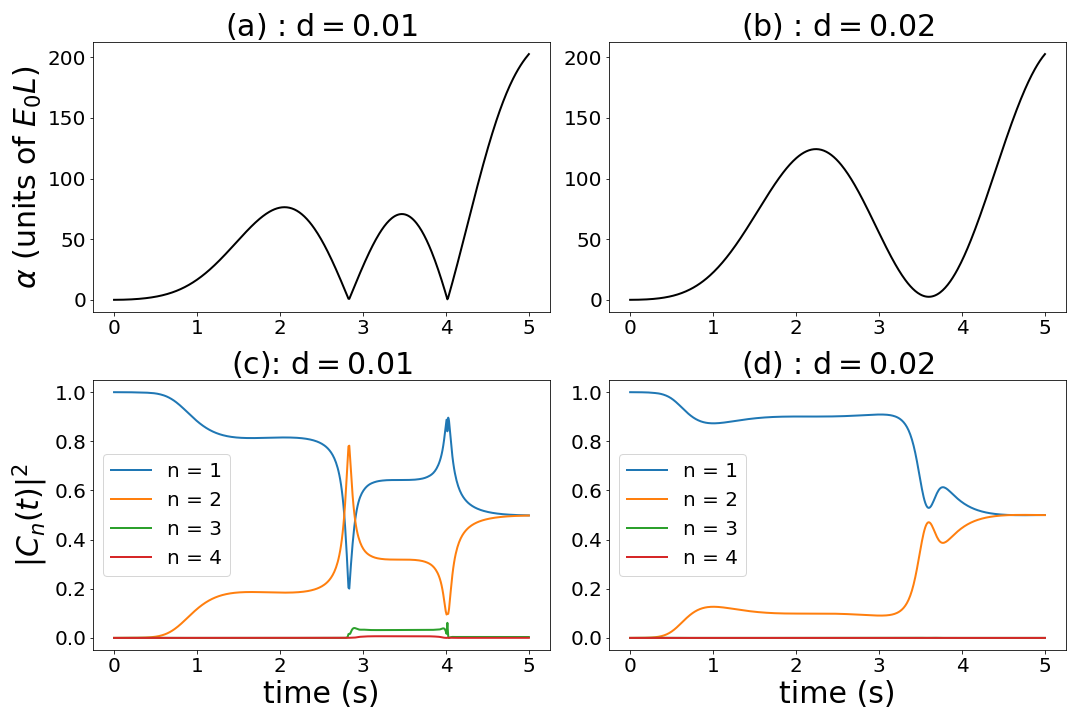}
	\caption{Results from the CRAB optimization for $ d = 0.01 $ and $ d = 0.02 $. In (a) and (b) we show the protocols $ \alpha(t) $, while in (c) and (d) we show the time evolution of $ \abs{c_n(t)}^2 $. We see that the protocol in (b) gives negligible excitations to states $ n>2 $ throughout its duration. However the protocol in (a) excites the third eigenstate during the first discontinuity in $ \dot{\alpha}(t) $ right before $ t= 3 $, but this excitation is depleted during the second discontinuity around $ t = 4 $.}
	\label{fig:CRAB_51_and_52}
\end{figure}
We use chopped random-basis (CRAB) optimization \cite{PhysRevA.84.022326} to find protocols $ \alpha(t) $ that splits the wave function in two equal halves for asymmetric barrier insertion in a quantum box.
In CRAB optimization we expand the protocol in a complete basis (the Fourier series in our case), in the following way
\begin{equation}
\alpha(t) = \alpha_0(t)\left[1 + \lambda(t)\sum_{n=1}^{N_c}A_n\cos(\omega_n) + B_n\sin(\omega_n)\right].
\end{equation}
Here $ \alpha_0(t) $ is an initial guess for the optimal protocol, $ \lambda(t) $ is a regularization function used to implement boundary conditions, and $ \left\lbrace A_n,B_n,\omega_n\right\rbrace  $ is the set of Fourier coefficients we optimize to maximize the cost function
\begin{equation}\label{cost_function}
C(\left\lbrace A_n,B_n,\omega_n\right\rbrace ) = 1 -  \sum_{n=1}^{2}\left(\abs{c_n(T)}^2-0.5\right)^2.
\end{equation}
We fix the boundary conditions to be $ \alpha(0) = 0 $ and $ \alpha(T) = 200 E_0 L $ (where $ E_0 = \pi^2/2 $ is the ground state at $  \alpha(0)=0 $) , and choose $ \lambda(t) = \sin(\pi t/T) $. To minimize Eq.~(\ref{cost_function}) we use a gradient free method, like the Nelder-Mead \cite{nelder1965simplex} or Powell's method \cite{powell1964efficient}. Using the Nelder-Mead method we are able to almost exactly split the wave function in half, and results for $ d = 0.01 $ and $ d = 0.02 $ are shown in Fig.~\ref{fig:CRAB_51_and_52}. In these examples we obtained $ \abs{c_1(T)}^2 = 0.4986 $, $ \abs{c_2(T)}^2 = 0.4979 $, and $ \sum_{n>2} \abs{c_n(T)} \simeq 10^{-3} $ for $ d = 0.01 $, while for $ d = 0.02 $ we got $ \abs{c_1(T)}^2 = 0.5001 $, $ \abs{c_2(T)}^2 = 0.4999 $, and $ \sum_{n>2} \abs{c_n(T)}^2 \simeq 10^{-5} $. In Fig.~\ref{fig:CRAB_51_and_52}(a) and Fig.~\ref{fig:CRAB_51_and_52}(b) we show example protocols for $ d = 0.01 $ and $ d = 0.02 $, respectively, while in Fig.~\ref{fig:CRAB_51_and_52}(c) and Fig.~\ref{fig:CRAB_51_and_52}(d) we show the time evolution of the probability to be in a given eigenstate, $ \abs{c_n(t)}^2  $.\\

\begin{figure}[h!]
\centering
\includegraphics[width=1\linewidth]{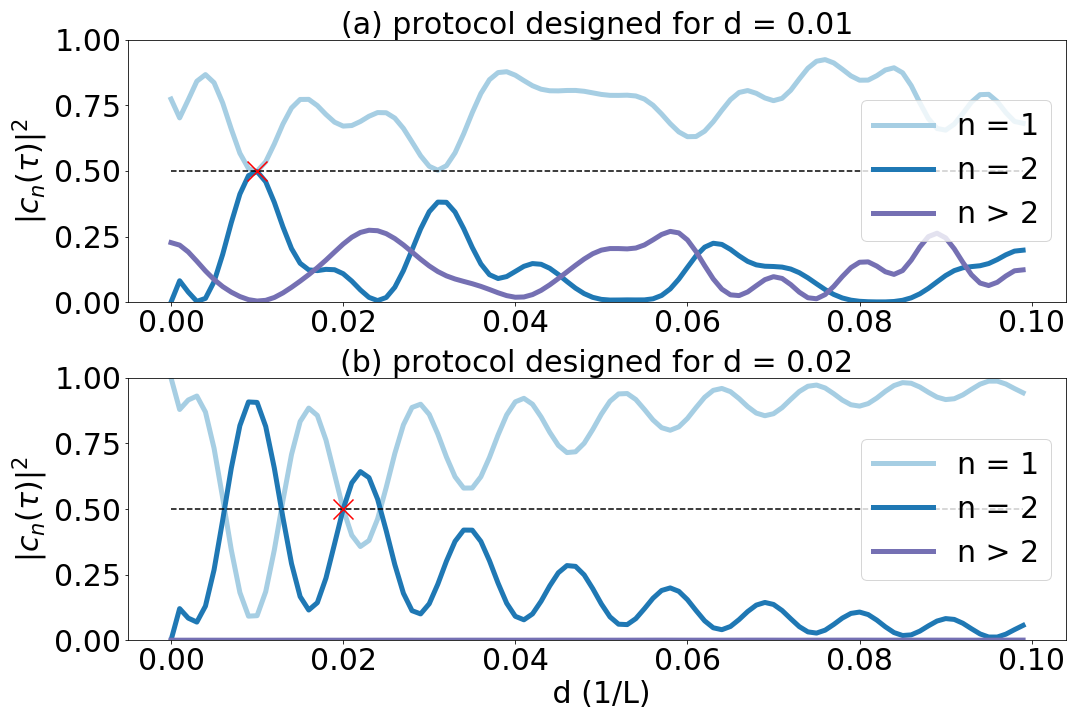}
\caption{Plot showing how the protocols designed for two specific asymmetries performs on other asymmetries. In (a) we show the results for the protocol designed for $ d = 0.01 $, while in (b) we show the one designed for $ d = 0.02 $. The light blue and the blue line shows the occupation at $ t = T $ for the first and second eigenstate respectively, while the purple line shows the occupation of all eigenstates higher than the second, i.e. the unwanted excitations. The black dashed lines shows the target $ \abs{c_n(T)}^2 = 0.5 $, and the red crosses shows the asymmetry trained on.}
\label{fig:CRAB_all_asym}
\end{figure}

The protocols obtained by CRAB are designed to split the wave function in two for a given asymmetry. They work extremely well for the asymmetry they were designed for. However, the protocols generalize poorly to other asymmetries, as shown in Fig.~\ref{fig:CRAB_all_asym}. There we plot $ \abs{c_n(T)}^2 $ as a function of the asymmetry $ d $, using the protocol designed for $ d = 0.01 $ and $ d = 0.02 $. We see that the performance of a protocol designed for a specific asymmetry dramatically reduces if it is applied to single-particle-boxes of different asymmetries. An interesting feature is seen in Fig.~\ref{fig:CRAB_all_asym}(b), where the protocol designed for $ d = 0.02 $ achieves exact splitting for asymmetries other than the one that was used for training. However, even this protocol has bad performance in the regions between these points of exact splitting, so it would not be useful unless one knows the exact asymmetry of the single-particle-box.

\section{Deep Q-Learning}
\begin{figure}
	\centering
	\includegraphics[width=1\linewidth]{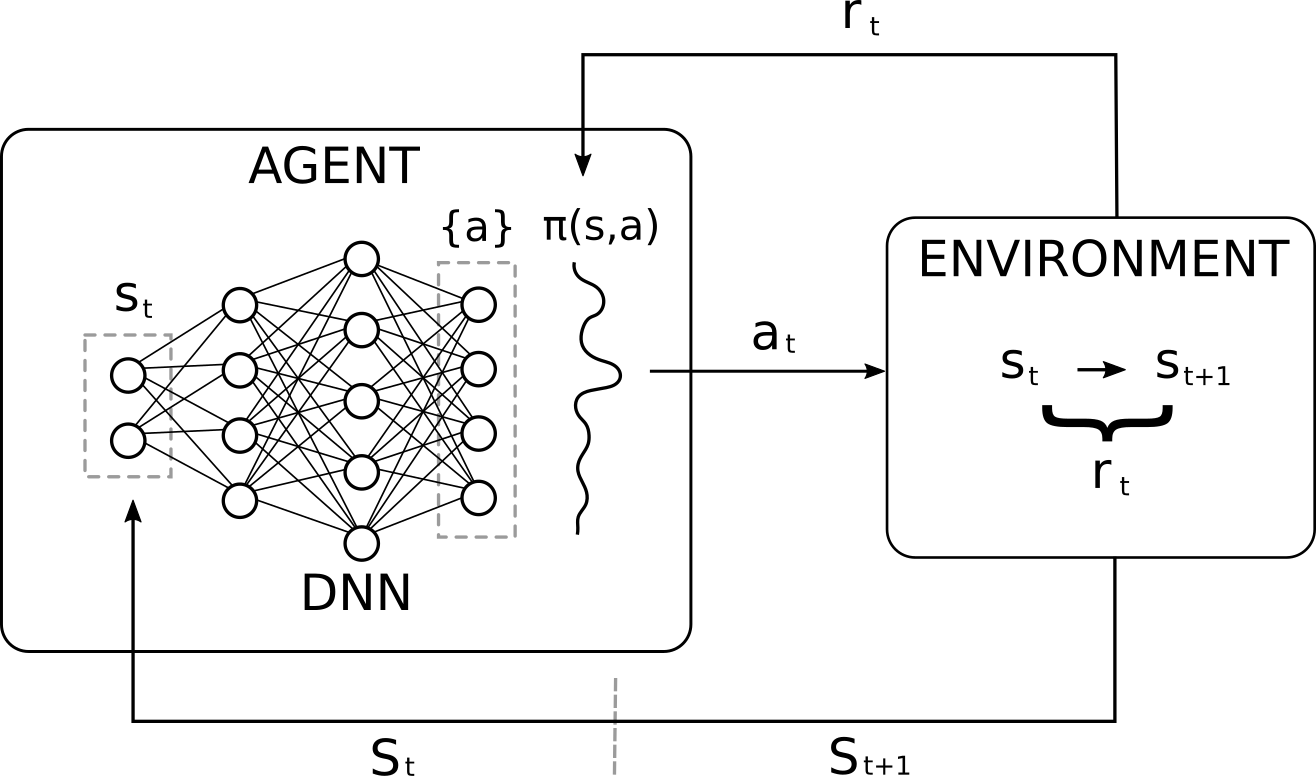}
	\caption{Schematic showing the basic setup of deep Q-learning. The current state of the system $ s_t $ is fed as input nodes into a deep neural network (DNN). The output nodes are the set of all possible actions, $ \set{a} $, and their values are the estimated Q-value for the given state-action pair. The policy $ \pi(s,a) $ is given by the action node with the highest output value, or by a random action if the agent is exploring. The action determined by the policy is performed in the environment, inducing a state change from $ s_t\to s_{t+1} $. Associated with this state change, a reward $ r_t $ is given, which is used to determine how good the given action was in this state. This reward is fed back into the DNN and used to update its weights according to Eq.~(\ref{eq:DQN_loss}). Schematic adapted from \cite{mao2016resource}.}
	\label{fig:DRL_basic}
\end{figure}
We now give a short review of the DQL algorithm introduced in \cite{mnih2015humanlevel}. In the next section we will show how this general algorithm is adapted to our problem. A schematic of the basic reinforcement learning protocol is shown in Fig.~\ref{fig:DRL_basic}. At time $ t $ the environment is in a given state $ s_t $. The agent performs an action $ a_t $ which induces a state change of the environment from $ s_t $ to $ s_{t+1} $. The agent then receives an observation of the new state of the environment, $ s_{t+1} $. After taking an action the agent receives a reward $ r_t = r(s_t,a_t,s_{t+1}) $. The reward function $ r(s_t,a_t,s_{t+1}) $ is designed by us, according to what goal we want the agent to achieve. \\

The behavior of the agent is determined by its policy $ \pi(a_t|s_t) $, which is the probability of taking the action $ a_t $ in given the observation $ s_t $. If the agent is in state $ s_t $, the Q-function (quality function) $ Q_\pi(s_t,a_t) $ gives the expected cumulative reward given that the action $ a_t $ is performed and the policy $ \pi $ is followed for all proceeding states.
\begin{eqnarray}
\nonumber Q(s_t,a_t) &=& \text{E}_{s_{t+1}} \left[r_t + \gamma r_{t+1} + \gamma^2r_{t+2}+\dots|s_t,a_t,\pi\right]\\
&=& \text{E}_{s_{t+1}}\left[r_t + \gamma Q(s_{t+1},a_{t+1}) | s_t,a_t,\pi\right]
\end{eqnarray}  
Here $ \gamma \leq 1$ is a discount parameter, which determines how much the agent values immediate reward compared to future reward. If $ \gamma < 1 $ the agent will value future reward less than immediate reward, which is useful for learning in stochastic environments where the future is more uncertain. The optimal Q-function, $ Q_\pi^*(s_t,a_t) $, is the maximum expected cumulative reward obtained by taking the action $ a_t $ in state $ s_t $ and then acting optimally thereafter, and it is shown to obey the Bellman optimality equation \cite{bellman1957markovian}
\begin{equation}\label{Bellman}
Q_\pi^*(s_t,a_t) = \text{E}_{s_{t+1}}\left[r_t + \gamma\max_{a_{t+1}} Q_\pi^*(s_{t+1},a_{t+1}) | s_t,a_t\right]
\end{equation} 
If we have $ Q^*_\pi(s,a) $ for all possible state-action pairs, it is clear that we can find the optimal policy, $ \pi^* $, by choosing $ a_t = \arg\max_{a'} Q_\pi^*(s_t,a') $, i.e. following the policy\\
\begin{equation}
\pi^*(a_t|s_t) = \underset{a'}{\arg\max}~ Q_\pi^*(s_t,a').
\end{equation}
The key idea introduced in \cite{mnih2015humanlevel}, is to estimate the optimal Q-function using a neural network $ Q_\pi^*(s,a)\simeq Q_\pi^*(s,a,\theta) $, where $ \theta $ is the weights and biases of the neural network. This neural network is called a Deep-Q network (DQN), and is updated by performing gradient ascent on the mean-squared-error of the current predicted $ Q_\pi^*(s,a,\theta) $, while using the Bellman equation as the target. The loss function for DQN is therefore\\
\begin{equation}\label{eq:DQN_loss}
L(\theta) = E_{s_{t+1}}\left[\big(Q_\pi^*(s_t,a_t,\theta) - y_t\big)^2\right].
\end{equation}
where
\begin{equation}\label{eq:DQN_y}
y_t = r_t + \gamma ~\underset{a_{t+1}}{\text{max}}~ Q_\pi^*(s_{t+1},a_{t+1},\theta)
\end{equation}
To create the neural network we used tensorflow's implementation of the Keras API \cite{tensorflow2015-whitepaper,chollet2015keras}, with Adam \cite{kingma2014adam} as the optimizer. The network consists of three hidden layers, with 24, 48 and 24 neurons, respectively, as well as 2 input neurons and 20 output neurons.
When the network is initialized its predictions for the optimal  $Q_\pi^*$-values are of course totally wrong. So if we always chose the actions that maximizes the current predicted $Q_\pi^*$-values, the agent would not learn anything. We need to let the agent explore the state-action space by randomly performing actions. A typical exploration policy is the $ \epsilon $-greedy policy. The agent chooses random actions with probability $ \epsilon $, or the ones with the highest $Q_\pi^*$-value (greedily) with probability $ 1-\epsilon $. As time goes and the agent explores more of the environment, $ \epsilon $ is decreased so that it focuses more on the areas of the state-action space with higher $Q_\pi^*$-values by taking deterministic actions. Typically we start by taking completely random actions, $ \epsilon = 1 $, and let $ \epsilon $ converge to some finite number $ \epsilon\sim 0.05 $, so that there is always some exploration going on. As seen in Eq.~(\ref{eq:DQN_loss}) a single update of the network weights requires the following input: the current state $ s_t $, the action chosen $ a_t $, the immediate reward $ r_t $, and the next state $ s_{t+1} $. We call this tuple, $ e_t = (s_t,a_t,r_t,s_{t+1}) $, that the network trains on an experience. Instead of training on consecutive experiences we store them all in a memory $ M_N = \left\lbrace e_0,e_1,\dots,e_N \right\rbrace  $, and then train on randomly drawn batches of samples from the memory. The memory have a finite capacity, and new experiences replace older ones when the memory is full. There are three main advantages of training on the  memory: It is data efficient since a single experience can be drawn many times. Only training on consecutive experiences is inefficient, since the network tends to forget previous experiences by overwriting them with new experiences. The time-correlation of consecutive experiences means that the network update due to the current experience determines what the next experience will be, so training can be dominated by experiences from a certain area in the state-action space. Finally we see that in Eq.~(\ref{eq:DQN_loss}) the current weights of the network determines both the target $Q_\pi^*$-value and the predicted $Q_\pi^*$-value from the Bellman equation. Thus every network update changes the target $Q_\pi^*$-value that we are trying to reach, and makes it hard for the network weights to converge. A simple way to circumvent this problem is to use two neural networks, one for the target $ Q_\pi^* $-value ($ \theta^- $), and one for the current $ Q_\pi^* $-value ($ \theta $). The target network is softly updated during training according to $ \theta^- \leftarrow \theta^-(1-\tau) + \theta\tau $, where $ \tau $ is a hyper-parameter that determines how close the two networks are in the network parameter space.

\section{DQL results}
For our system, we defined the state to be a tuple of the strength of the $ \delta $-barrier and the time $ t $, i.e. $ S = \lbrace\alpha(t),t\rbrace $. The available actions is a set of $ \dot{\alpha}(t) $, given by 
\[ A = \set{\dot{\alpha}_{n}^\pm (t) = \pm~ 2^n, \quad\text{for}\quad n = 1,2,\dots,10 } .\]
The initial state is $ S = \lbrace\alpha(t) = 0, t = 0\rbrace $, and the goal is to reach a state where $ \abs{c_0}^2 =\abs{c_1}^2 = 1/2 $, at the end of the protocol $ t = T $. A sequence of selected actions, from time $ t=0 $ to $ t = T $, defines a protocol $ \alpha(t) $. The number of times the agent chooses an action per protocol is given by $ N_t $, and the time-step is therefore $ dt = T/N_t $. The environment that the agent acts in is the quantum mechanical SPB, with initial state $ \abs{c_1}^2 = 1 $ and $ \abs{c_{n>1}}^2 = 0 $, and time evolution given by the Schr\"{o}dinger equation $ i\partial_t\ket{\psi(t)} = H(t)\ket{\psi(t)}$, which we solve as in \cite{PhysRevA.99.022121}. The sequential process for one episode is then 
\begin{enumerate}
	\item Initial state is $ s_0 : (\alpha_0 = 0,t = 0) $
	\item Agent chooses action based on $ s_0 $, e.g. $ a_0 = \dot{\alpha}_3^+ $
	\item The next state is then $ s_1: (\alpha_0 + \dot{\alpha}_3^+dt, t + dt) $
	\item Repeat 2. $ \to $ 3., for $ s_1,s_2,\dots $ until $ t = T $.
	\item Solve the Schr\"{o}dinger equation for the given protocol (set of all states $\set{s_n,t_n}$) and calculate reward. Repeat from 1. until maximum number of episodes reached.
\end{enumerate}
The reward function we used is defined by
\begin{equation}
r(t)=\begin{cases}
0,\qquad    \text{ if } t<T \text{ and } \alpha\in[0,\alpha_{max}]\\
-10,\quad  \text{if } t<T \text{ and } \alpha\notin[0,\alpha_{max}] \\
100\exp(-\sum_{n=1}^{2}\frac{\left(\abs{c_n(T)}^2-0.5\right)^2}{\sigma}), \quad \text{if } t = T
\end{cases}
\end{equation}
where $ \sigma $ determines how sharp we want the reward distribution to be. If the agent chooses actions such that $ \alpha(t) < 0 $, we give it a punishment of $ - 10 $ and set $ \alpha(t) = 0 $, and for actions that would give $ \alpha(t) >  \alpha_{max} $ we punish and set $ \alpha(t) = \alpha_{max} $. We do this to keep the state space bounded. The space of possible protocols grows exponentially with $ dt^{-1} $, so it is impractical to set $ dt $ so small that we get approximately continuous $ \dot{\alpha}(t) $. The accuracy of our numerical solution of the quantum time evolution decreases if we have discontinuous $ \dot{\alpha}(t) $, so to circumvent this problem we use cubic spline to interpolate the final protocol before calculating the reward.
\begin{figure}
\centering
\includegraphics[width=1\linewidth]{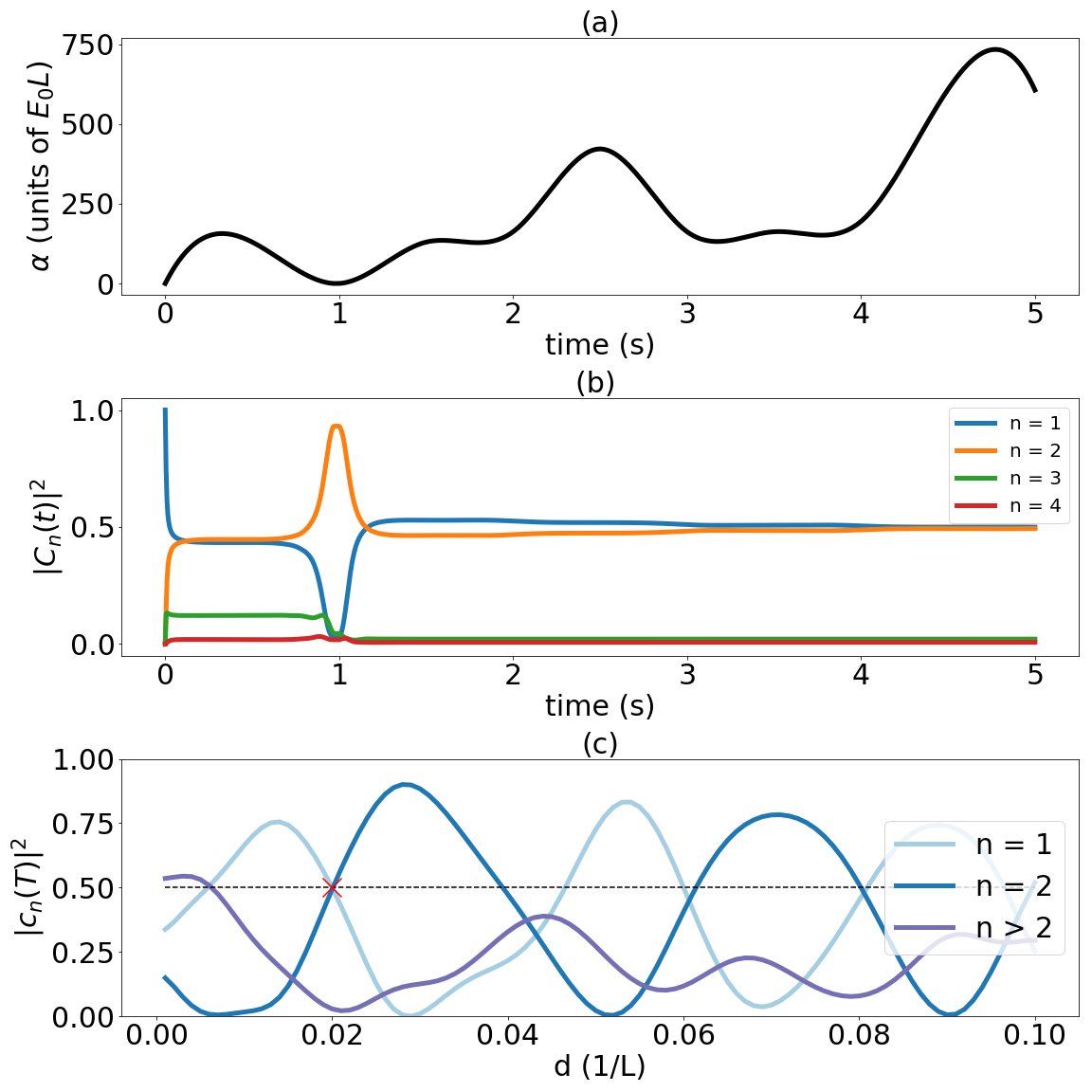}
\caption{Results from DQL when training on a single asymmetry, $ \epsilon = 0.02 $. In (a) we show the protocol $ \alpha(t) $, while in (b) we show the time evolution of $ \abs{c_n(t)}^2 $ for the asymmetry we trained on. Similarly to the protocol in Fig.~\ref{fig:CRAB_51_and_52}(a), there is a good amount of excitations to the third eigenstate in the very beginning of the protocol, which is then depleted around $ t = 1 ~s $. In (c) we show how the protocol generalizes to other asymmetries, by plotting the distribution $ \abs{c_n(T)}^2 $ at $ t = T $ for asymmetries in the range $ d \in[0.01,0.1] $. The parameters of this protocol was $ T = 5~s $, $ N_t = 10 $, $ \alpha_{max} = 800~E_0L $, and $ \sigma = 0.05 $.}
\label{fig:DQL_52}
\end{figure}\\
In Fig.~\ref{fig:DQL_52} we show an example protocol learned by the DQL agent, and the corresponding time evolution of $ \abs{c_n(t)}^2 $, when training on a single asymmetry ($ \epsilon = 0.02 $) for 10 000 episodes. The final distribution was $ \abs{c_1(T)}^2 = 0.4996 $, $ \abs{c_2(T)}^2 = 0.4935 $, and with higher excitations  $ \sum_{n>2} \abs{c_n}^2 \simeq 10^{-2} $. The results, when training on a single asymmetry, tended to be worse for DQL than for direct CRAB optimization. There are many ways to improve the results obtained by DQL; we can add actions to, or change the action space, train for a longer time or increase the number of actions per episode $ N_t $. Alternatively, one could implement algorithms similar to DQL that can perform actions in a continuous action space, like deep deterministic policy gradient (DDPG) \cite{lillicrap2015continuous}. However, most of these changes would also increase the necessary training time. \\

In real experiments one may not know exactly how large the asymmetry of the single-particle-box is. A far more useful protocol would be a robust one, designed to work best for a given range of asymmetries. One of the main benefits of DQL is that it is a model-free algorithm, so this task is easily achieved. One only has to let the agent train on random samples of the set of asymmetries one wants the protocol to be optimized to. Since the agent tries to maximize the expected cumulative reward, this added stochasticity is no hindrance. How much the agent values a given state-action pair is averaged over the random samples from the memory, which is proportionally filled with the number of asymmetries we train on.\\
\begin{figure}
\centering
\includegraphics[width=1\linewidth]{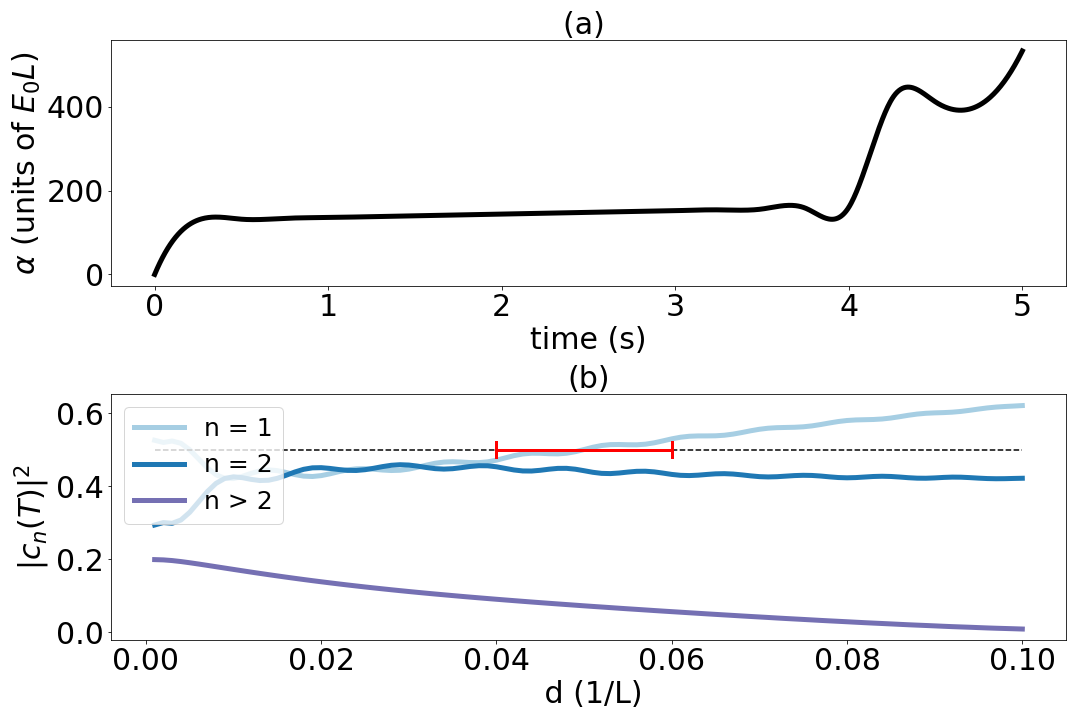}
\caption{Results from DQL when training on 10 different asymmetries in the range $ d \in\left[0.04,0.06\right] $. In (a) we show the protocol obtained, while in (b) we show $ \abs{c_n(T)}^2 $ all asymmetries up to $ d = 0.1 $, where the red bar indicates the range of asymmetries we trained on. When compared to Fig.~\ref{fig:CRAB_all_asym}, we see that the protocol performs much better overall than the ones designed for one specific asymmetry, particularly in the range we trained on. The parameters of this protocol was $ T = 5~s $, $ N_t = 20 $, $ \alpha_{max} = 800~E_0L $, and $ \sigma = 0.05 $.}
\label{fig:DQL_all_asymmetries}
\end{figure}
\begin{figure}
\centering
\includegraphics[width=1\linewidth]{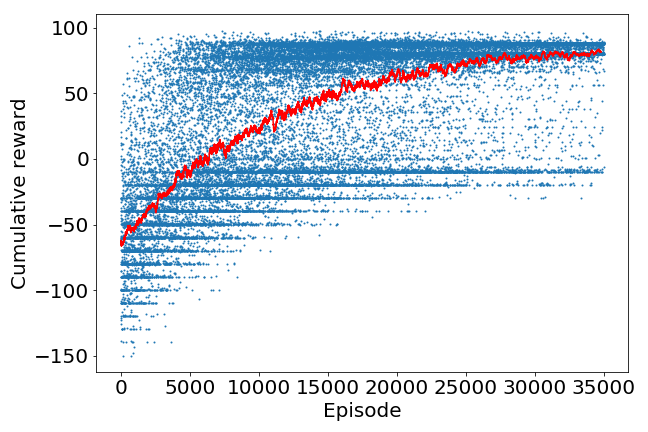}
\caption{Scatter plot of the reward received per episode, when training on multiple asymmetries, shown in blue dots, and a running average shown in red. The probability to take random actions is gradually reduced with the number of episodes, leading to a final protocol which the agent determines to be the best.}
\label{fig:DQL_reward_all_asymmetries}
\end{figure}
As an example, say one could determine the asymmetry with a given accuracy $ d = 0.05 \pm 0.01 $. An example protocol that was obtained when training on multiple asymmetries (10 equally spaced samples in the range $ d \in \left[0.04,0.06\right]$) is shown in Fig.~\ref{fig:DQL_all_asymmetries}. As seen in Fig.~\ref{fig:DQL_all_asymmetries}(b), this protocol performs better on the full range of asymmetries than the ones designed for a single asymmetry, shown in Fig.~\ref{fig:CRAB_all_asym}. The excitation to states higher than the two first eigenstates is largest for small asymmetries. This is due to the fact that when $ d\to 0 $, the wall is inserted close to the central node of the second eigenstate, and the central anti-node of the third eigenstate, as shown in Fig.~\ref{fig:spb}. Therefore excitations to the second eigenstate becomes less likely, while the opposite is true for excitations to the third eigenstate. Since this is an intrinsic property of the system, it is impossible to find protocols that avoids excitations for $ d\to 0 $. For $ d=0 $, the ground state of the left and right compartment constitute a doubly degenerate global ground state, and to achieve an equal splitting of the wave function, one has to insert the barrier adiabatically \cite{PhysRevA.99.022121}. \\

In Fig.~(\ref{fig:DQL_reward_all_asymmetries}) we see the total cumulative reward received per episode in a scatter plot, as well as a running average. We see that in the early episodes, where there agent mostly performs random actions, there are many episodes with negative cumulative reward. This is because there is an equal probability that the agent chooses negative and positive $ \dot{\alpha} $, and since the initial state is $ \alpha(t = 0 ) = 0 $ there is a high probability that the agent chooses actions which gives $ \alpha(t)<0 $, resulting in a punishment of -10 every time. In this early stage the agent explores and learns about its environment. As the probability to take random actions decreases (according to the $ \epsilon $-greedy protocol) with each episode, the agent takes more deterministic actions based on its experience, and the reward per episode increases steadily. The stochasticity observed in the rewards for final episodes is due a finite final exploration rate $ \epsilon = 0.05 $. We obtain the final protocol after training by setting $ \epsilon = 0 $, and let the agent act deterministically. The efficiency of the protocol obtained by training on a range of asymmetries can be increased by implementing the same changes as for the one designed for a single asymmetry.\\

\section{Deep Deterministic Policy Gradient}
Our set of possible actions for the DQL algorithm is somewhat arbitrarily chosen. For our specific control problem, there are infinitely many protocols that can achieve our goal, so the exact set chosen is not critically important. However,  the performance of the algorithm  depends on this choice, and the optimal protocols we find can always be defined by some subset of the total action-space. That is, not all actions are used for the optimal protocol, so we could retroactively reduce the action-space after learning which actions was needed. For many control problems in physics, it is more natural to let the action values be drawn from a continuous set, on some interval $ A \in [a_{min},a_{max}] $. For DQL, this is not possible, since the optimal policy $ \pi^*(a_t|s_t) $ comes from taking the maximum argument of a finite dimensional $ Q^*(s_t,a_t) $.\\

When the action-space is continuous, the optimal Q-function $ Q^*(s,a) $ is assumed to be differentiable with respect to the action $ a $.
In Deep Deterministic Policy Gradient \cite{lillicrap2015continuous}, the goal is to find a deterministic policy $ \mu(s) $, which gives is the optimal action to take for any state, $ a^* = \mu(s) $. This deterministic policy is approximated by another neural network $ \mu(s)\simeq \mu(s,\phi) $, where $ \phi $ are the parameters of the network. The Q-function is, as in DQL, also approximated by a neural network, and the essence of introducing the deterministic policy is to replace the largest Q-value for a state-action pair in the following way:
\begin{equation}
\arg\max_{a'}~ Q_\pi^*(s_t,a',\theta)\to Q^*(s_{t+1},\mu(s_{t+1},\phi),\theta).
\end{equation}
The Q-network is updated in the same way as in DQL, by using the Bellman equation, but instead of Eq.~(\ref{eq:DQN_y}), the target for the loss function now becomes
\begin{equation}
y_t = r_t + \gamma Q^*(s_{t+1},\mu(s_{t+1},\phi),\theta).
\end{equation}
As for the policy network, it was shown in \cite{silver2014deterministic} that its weights can updated in proportion to the gradient of the Q-function
\begin{equation}\label{eq:policy_update}
\phi_{k+1} = \phi_k + \lambda E_{s\in B}\left[\grad_\phi{Q^*(s,\mu(s,\phi),\theta)}\right],
\end{equation}
where $ \lambda $ is the learning rate, which determines the step-size of the gradient ascent. Since the gradient will, in general, move the weights in different directions for different states, an average over a batch of experiences is taken. By applying the chain rule to Eq.~\ref{eq:policy_update}, we can decompose it into a product of the gradient of the policy with respect to its network weight, and the gradient of the Q-function with respect to the actions
\begin{equation}
\grad_\phi{Q^*(s,\mu(s,\phi),\theta)} = \grad_\phi{\mu(s,\phi)}\grad_aQ^*(s,a,\phi)|_{a = \mu(s,\phi)}.
\end{equation}
Exploration in DDPG is driven by adding noise to the policy, sampled form some distribution $ N_t $ suited to the environment, which is annealed over time
\begin{equation}
\mu'(s_t) = \mu(s_t,\phi) + N_t.
\end{equation}
We use a Gaussian white noise process, and annealed its standard deviation from $ \sigma_N = 0.3 $ to $ \sigma_N = 10^{-4} $ over the course of the training.
DPPG is called an actor-critic model, and the sense is that the policy is an actor, taking actions in an environment, and the Q-function acts as a critic, determining how good the actions where, and feeding the result back to the actor.\\

For the DPPG algorithm, we used an adapted implementation from Keras-RL \cite{plappert2016kerasrl}, which includes the same modifications we used for DQL; i.e. experience replay and different networks for the target and current Q-function and policy. The policy network takes as input the same state tuple as for DQL, $ S=\set{\alpha(t),t} $, which is connected to three hidden layers, with the same architecture as DQL; 24, 48, and 24 neurons, respectively, and outputs a single value, the action $ \dot{\alpha}(t) $. The Q-function network takes as input the action value suggested by the policy, as well as the state $ S = \set{\alpha(t),t} $, again connected to three hidden layers with the same architecture as DQL, and outputs a value which is its estimation of the optimal Q-value of the state-action pair. The output actions form the policy network are clipped at $ |\dot{\alpha}(t)|\leq 1000~E_0L/s $, and we use the same reward function as for DQL.\\

\begin{figure}
\centering
\includegraphics[width=1\linewidth]{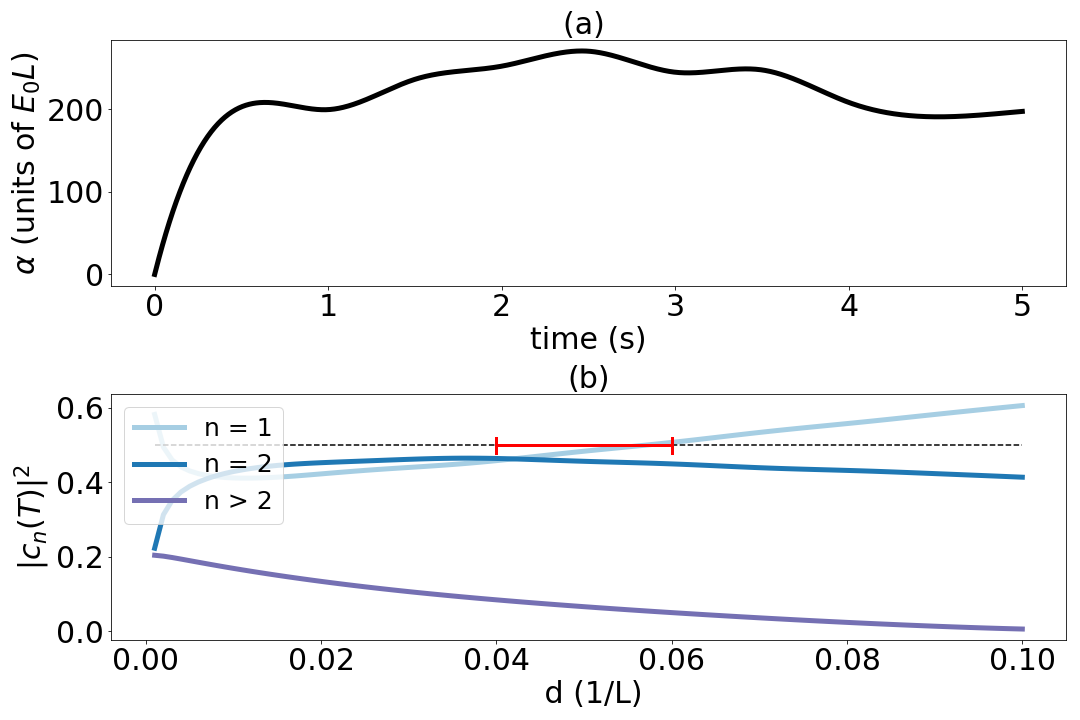}
\caption{Results from DDPG, when training on 10 different asymmetries in the range $ d\in[0.04,0.06] $. In (a) we show the protocol itself, while in (b) we show how the protocol performs on a range of asymmetries $d\in[0,0.1]  $. The red bar marks the range we trained on. The parameters of this protocol was $ T = 5~s $, $ N_t = 20 $, $ \alpha_{max} = 800~E_0L $, and $ \sigma = 0.05 $, and we trained for 20 000 episodes.}
\label{fig:DDPG_big_interval}
\end{figure}

In Fig.~\ref{fig:DDPG_big_interval}(a) we show a protocol obtained from DDPG, when training on 10 asymmetries in the range $ d \in [0.04,0.06] $, and in Fig.~\ref{fig:DDPG_big_interval}(b) the performance of the protocol on a range of asymmetries from $ d \in [0,0.1] $. As expected, the best results are obtained for the range of asymmetries we trained on, indicated by a red bar. A rigorous comparison between DQL and DPPG is difficult, partly due to the large amount of hyper-parameter tweaking needed to optimize each algorithm, but largely due to the arbitrary choice of discrete action values for DQL: for our example problem, there is no natural set of available actions to choose. As mentioned earlier, the performance of DQL for our problem, depends on the set of actions chosen, and therefore a fair comparison of the algorithms is complicated. The choice between discrete and continuous-action algorithms, has to be taken based on the specific problem one wants to solve. For our SPB problem, there are infinitely many "good" solutions, and since we interpolate the protocol at the end of each episode, both DQL and DPPG are well suited.\\

We used a 3.40 GHz CPU, and the training time for the most resource-intensive computation (the protocol in Fig.~(\ref{fig:DQL_all_asymmetries})) was about 48 hours, so increased training time is something that more advanced computation systems can handle. The most computationally-intensive part of the training, by a large margin, was solving the Schr\"{o}dinger equation after each episode. As for the hyper-parameters of the neural networks, we used a learning rate $ \lambda = 10^{-3} $, target network update every $ \tau = 10^{-3} $ time-step, and a replay memory size between $ 10\%-50\% $ of the total number of experiences. The $ \varepsilon $-greedy exploration policy was a linear decrease from $ \varepsilon = 1 $ to $ \varepsilon = 0.05 $.

\section{Discussion and Summary}
We have used CRAB optimization and deep reinforcement learning to construct protocols, $ \alpha(t) $, for the time-dependent strength of a barrier inserted asymmetrically in a single-particle-box, in such a way that the wave function is split in two equal halves. These results implies that the asymmetric quantum Szilard engine can reach the same efficiency in information-to-work conversion as the symmetric one, since no information is lost in the which-side measurement.\\

Using CRAB optimization, the protocols we obtain performs very well for the specific asymmetry we optimize for, but the protocol generalize poorly for different asymmetries. Although more time consuming and than CRAB optimization, we can also use DRL to find high performing protocols when training on single asymmetries. However, one of the biggest strengths of reinforcement learning based techniques is the possibility to perform robust and noise-resistant optimization. When training on a range of different asymmetries simultaneously, DRL can be used to find the protocols that performs best on the average of all the asymmetries sampled. Both DQL and DDPG were able to find good protocols for our example SPB problem, but in general, the choice between discrete and continuous-action algorithms has to be made on the basis of what specific problem one wants to solve. The advantage of using reinforcement learning for quantum control, is multifaceted: having model-free algorithms makes it simple to change the optimization criterion to make the agent solve different problems within the same environment, one only have to change the reward function to suit the new goal. Furthermore, since the agent is not tailored to any specific environment, it can easily be adopted to work in entirely different systems (e.g. we can use the agents constructed here to perform state-transfer in qubit systems \cite{bukov2018reinforcement}). Finally, the stochastic nature of the agents learning procedure is advantageous when one wants to perform robust optimization which can perform well with noise. These points all suggests that reinforcement learning can become a very useful tool in physics.\\

\section*{Acknowledgements}
We thank Y. M. Galperin for reading the manuscript and making valuable comments.

\end{document}